\documentclass{IEEEtran}

\usepackage[hidelinks]{hyperref}
\usepackage{caption}
\usepackage{indentfirst}

\usepackage{xcolor}
\usepackage{graphicx}
\usepackage{epstopdf}
\graphicspath{{fig/}}
\usepackage{subfigure}
\usepackage{float}
\usepackage{array}
\usepackage{url}
\usepackage{longtable}
\usepackage{rotating}
\usepackage{multirow}
\usepackage{amsmath}
\usepackage{amsfonts}
\usepackage{amssymb}
\usepackage{mathrsfs}
\usepackage{mdwlist}
\usepackage{booktabs}
\usepackage{threeparttable}
\usepackage{makecell}
\usepackage{diagbox}
\usepackage{pifont}
\usepackage{xcolor}
\usepackage{booktabs}
\usepackage{adjustbox}

\usepackage{pdflscape}
\usepackage{footnote}

\usepackage{bm}

\usepackage[numbers,sort&compress]{natbib}
\usepackage[linesnumbered,boxed,ruled]{algorithm2e}
\usepackage[hang,flushmargin]{footmisc}
\usepackage{lipsum}

\usepackage[T1]{fontenc}
\usepackage[utf8]{inputenc}

\usepackage{color,xcolor}
\definecolor{myred}{RGB}{217,46,127}
\definecolor{mygreen}{RGB}{67,127,127}

\title{DualSpec: Text-to-spatial-audio Generation via Dual-Spectrogram Guided Diffusion Model}

\author{
    {Lei Zhao, Sizhou Chen, Linfeng Feng, Jichao Zhang, Xiao-Lei Zhang, \IEEEmembership{Senior Member, IEEE}, Chi Zhang and Xuelong Li, \IEEEmembership{Fellow, IEEE}}

}

\begin{document}
\maketitle

\begin{abstract}
Text-to-audio (TTA), which generates audio signals from textual descriptions, has received huge attention in recent years. However, recent works focused on text to monaural audio only. As we know, spatial audio provides more immersive auditory experience than monaural audio, e.g. in virtual reality. To address this issue, we propose a text-to-spatial-audio (TTSA) generation framework named {DualSpec. Specifically,} it first trains variational autoencoders (VAEs) for extracting the latent acoustic representations from sound event audio. Then, given text that describes sound events and event directions, the proposed method uses the encoder of a pretrained large language model to transform the text into text features. Finally, it trains a diffusion model from the latent acoustic representations and text features for the spatial audio generation. In the inference stage, only the text description is needed to generate spatial audio. Particularly, to improve the synthesis quality and azimuth accuracy of the spatial sound events simultaneously, we propose to use two kinds of acoustic features. One is the Mel spectrograms which is good for improving the synthesis quality, and the other is the short-time Fourier transform spectrograms which is good at improving the azimuth accuracy. We provide a pipeline of constructing spatial audio dataset with text prompts, for the training of the VAEs and diffusion model. We also introduce new spatial-aware evaluation metrics to quantify the azimuth errors of the generated spatial audio recordings. Experimental results demonstrate that the proposed method can generate spatial audio with high directional and event consistency.
\end{abstract}

\begin{IEEEkeywords}
    Text-to-spatial-audio, audio generation, latent diffusion model.
\end{IEEEkeywords}

\section{Introduction} \label{sec:introduction}

\IEEEPARstart{S}patial audio generation is essential for immersive extended reality (XR) environments, interactive entertainment systems, and dynamic media production. With the fast development of Artificial Intelligence Generated Content, there is {a challenging question} whether we could generate spatial audio from simply text that describes the sound events and spatial directions, known as text-to-spatial-audio (TTSA) generation. TTSA is a brand new research direction. It is rooted in the {active research area} of text-to-audio (TTA) generation, which is a task of creating monaural audio from text descriptions. Its core challenge is how to guarantee the synthesis quality and azimuth accuracy of the spatial sound events simultaneously. To address this issue, TTA and conventional spatial audio generation techniques are involved. We summarize the two kinds of techniques as follows.

On the TTA side, the challenging issue of TTA is how to generate any kinds of sound events, covering from natural environments to human speech, flexibly with guaranteed high quality. Some recent development is as follows. AudioLM \cite{borsos2023audiolm} utilizes the w2v-BERT model \cite{chung2021w2v} to extract semantic tokens from audio inputs. AudioLDM \cite{liu2023audioldm} generates text-conditioned audio using a latent diffusion model (LDM) \cite{rombach2022high}, where the diffusion process is guided by CLAP embeddings \cite{elizalde2023clap} and utilizes a variational autoencoder (VAE) \cite{kingma2013auto} to generate latent embeddings of the Mel spectrograms of the audio. {\cite{huang2023make} achieves significant improvements in semantic alignment and temporal consistency of generated audio by adopting a dual-text encoding architecture, a feed-forward Transformer-based diffusion denoiser and data augmentation strategies driven by large language models.} Tango \cite{ghosal2023tango} is built on AudioLDM \cite{liu2023audioldm}. It replaces CLAP with a fine-tuned large language model (LLM) FLAN-T5 \cite{chung2024scaling}. Tango2 \cite{majumder2024tango} further improved the performance of Tango by employing direct preference optimization \cite{rafailov2023direct} and alignment training. {Auffusion \cite{10731578} leverages the advantages of text-to-image diffusion models in terms of generation capability and cross-modal alignment, significantly improving the quality of audio generation as well as the matching accuracy between text and audio under limited data and computational resources.}


On the side of the spatial audio generation, it can be categorized into conventional approaches and deep learning based approaches. Conventional digital signal processing (DSP) techniques create spatial sound by integrating two key acoustic models: room impulse responses (RIR) \cite{brinkmann2019cross} and head-related transfer functions (HRTF) \cite{zotkin2004rendering, jianjun2015natural, brinkmann2019hutubs}.  RIR characterize how sound waves propagate in physical environments by capturing room-specific reflections and reverberation. HRTF represent directional sound cues through frequency-dependent filtering effects caused by the interactions between sound waves and human anatomical structures (e.g., ears, head, torso).

In recent years, deep-learning-based spatial audio generation methods have shown remarkable potential for spatial audio generation. {The work in \cite{gebru2021implicit} demonstrated that neural networks can implicitly learn HRTF characteristics from data. Meanwhile, the work in \cite{richard2021neural} developed a convolutional architecture with neural time warping for addressing temporal misalignments between monaural and spatial audio.} Building on these foundations, recent deep learning approaches \cite{richard2021neural, leng2022binauralgrad} have achieved significant advances in monaural-to-spatial conversion. Notably, these methods share core principles with conventional DSP techniques. That is, both of them map monaural inputs to spatial targets using directional parameters such as the direction-of-arrival (DOA) information or quaternion coordinates. Their key distinction lies in replacing handcrafted acoustic models with data-driven representations learned by DNNs.

 However, there is little work on TTSA which seems a new topic combining the above two independent research directions. To our knowledge, the most related work is AudioSpa \cite{feng2025audiospa}. It generates spatial audio from a given monaural sound event and text that describes the azimuth of the sound event, where FLAN-T5 is used to transform the text into text embeddings. However, this approach relies on monaural audio as a reference condition rather than generating binaural audio directly from a single text prompt.

Based on the analysis above, we introduce an innovative TTSA model called DualSpec. This model generates spatial audio directly from text descriptions, without reference audio. The contributions can be summarized as follows:

\begin{itemize}
    \item
    \textbf{We propose DualSpec, an innovative dual-spectrogram guided generation framwork for TTSA.}
    {DualSpec utilizes two types of acoustic features to simultaneously enhance the generation quality and azimuth accuracy of spatial sound events. One is the Mel spectrograms, which are beneficial for improving generation quality, and the other one is the short-time Fourier transform (STFT) spectrograms, which excel at boosting azimuth accuracy. These acoustic features are compressed into latent representations, and then fed into a diffusion model for training and inference.}

    \item
\textbf{We design multiple variational autoencoders (VAEs). They can efficiently compress different acoustic features into low-dimensional latent representations.} Specifically, these features include Mel spectrograms, STFT spectrograms, and their combinations. These latent features are used to train various diffusion models, providing a foundation for generating high-quality and location-controllable spatial audio.

    \item
    {\textbf{We present a pipeline for constructing spatial audio datasets.} We process the collected monaural audio with head-related impulse responses (HRIRs) convolution to generate binaural spatial audio and annotate a portion of the audio with text descriptions of the audio event and spatial information. The spatial audio dataset is used for self-supervised training of VAEs, while the subset of the binaural spatial audio with text descriptions is utilized for training the diffusion model.}

    \item
    \textbf{We employ spatial perception metrics to evaluate the directional accuracy of generated spatial audio.} The measurement utilizes a pre-trained sound source localization model, which accurately calculates the azimuth error between the generated audio and the location in text. This provides an objective and quantitative way to evaluate the spatial quality of the generated audio.

\end{itemize}

The rest of this paper is organized as follows. Section \ref{sec:re} presents some preliminaries. Section \ref{sec:DualSpec} describes the proposed method in detail. Section \ref{sec:local} describes the sound localization model. Section \ref{sec:data} explains the construction pipline of the spatial audio dataset. Section \ref{sec:exp} presents the experimental setup and results. Finally, Section \ref{sec:con} concludes the study.

\section{Preliminaries}   \label{sec:re}

Diffusion models are a class of generative models that learn data distributions through iterative noise injection and denoising processes. These models learn to reverse a gradual noising process through an iterative denoising procedure. Among various diffusion model variants, the denoising diffusion probabilistic model (DDPM) \cite{ho2020denoising, Song2021} has become particularly influential.

A DDPM operates over $T$ steps, with two key stages: forward diffusion and reverse generation, both modeled as a $T$-step Markov chain. In the forward diffusion, noise is added to the initial sample $x_0$ over $T$ steps, resulting in the noisy sample $x_T$. The reverse process aims to reconstruct $x_0$ from $x_T$. Due to the Markov property, each step $t$ depends on the previous step $t-1$, as expressed by:

\begin{equation}
q(x_1, \dots, x_T \mid x_0) = \prod_{t=1}^{T} q(x_t \mid x_{t-1}),
\end{equation}
where
$
q(x_t \mid x_{t-1}) = \mathcal{N}(x_t; \sqrt{1-\beta_t} x_{t-1}, \beta_t \mathbf{I}).
$
Here, $\beta_t$ typically increases with each step ($\beta_1 < \beta_2 < \dots < \beta_T$).

By reparameterization, the state at step $t$ can be expressed as:
\begin{equation}
x_t = \sqrt{\bar{\alpha}_t} x_0 + \sqrt{1-\bar{\alpha}_t} \bm{\epsilon},
\end{equation}
where $\bm{\epsilon} \sim \mathcal{N}(0, \mathbf{I})$, $\alpha_t = 1 - \beta_t$, and $\bar{\alpha}_t = \prod_{i=1}^t \alpha_i$. The distribution for $x_t$ given $x_0$ is:

\begin{equation}
q(x_t \mid x_0) = \mathcal{N}(x_t; \sqrt{\bar{\alpha}_t} x_0, \sqrt{1-\bar{\alpha}_t} \mathbf{I}).
\end{equation}

For the reverse process, to reconstruct $x_0$ from $x_T$, the conditional distributions are parameterized as:

\begin{equation}
p_{\theta}(x_0, \dots, x_{T-1} \mid x_T) = \prod_{t=1}^{T} p_{\theta}(x_{t-1} \mid x_t),
\end{equation}
with the reverse transition given by:
$
p_{\theta}(x_{t-1} \mid x_t) = \mathcal{N}(x_{t-1}; \mu_{\theta}(x_t, t), \sigma_{\theta}(x_t, t)^2 \mathbf{I}),
$
where
\begin{equation}
\mu_{\theta}(x_t, t) = \frac{1}{\sqrt{\alpha_t}} \left( x_t - \frac{\beta_t}{\sqrt{1 - \bar{\alpha}_t}} \epsilon_{\theta}(x_t, t) \right),
\end{equation}
and
\begin{equation}
\sigma_{\theta}(x_t, t)^2 = \tilde{\beta}_t.
\end{equation}

For $t > 1$, $\tilde{\beta}_t = \frac{1 - \bar{\alpha}_{t-1}}{1 - \bar{\alpha}_t} \beta_t$, and for $t = 1$, $\tilde{\beta}_1 = \beta_1$. $\epsilon_{\theta}(x_t, t)$ is a neural network estimating the noise $\bm{\epsilon}$ at time $t$.

The model is trained by minimizing the objective:

\begin{equation}
\mathbb{E}_{x_0, t, \epsilon}\left\|\epsilon - \epsilon_{\theta}\left(\sqrt{\bar{\alpha}_t} x_0 + \sqrt{1 - \bar{\alpha}_t} \bm{\epsilon}, t\right)\right\|_2^2,
\end{equation}
where $t$ is chosen randomly from $\{1, \dots, T\}$. The network $\epsilon_{\theta}$ is optimized to estimate $\bm{\epsilon}$, aiding in the recovery of the original input. During inference, the reverse process starts with a random sample $x_T \sim \mathcal{N}(0, \mathbf{I})$, and iteratively refines it back to $x_0$ using the learned model.

However, applying diffusion models directly to high-dimensional data like images poses computational challenges and requires significant memory resources. {To address these limitations, LDM transfers the diffusion process from a high-dimensional data space to a lower-dimensional latent space.} This approach was first systematically applied to image generation in the work Stable Diffusion \cite{rombach2022high}. In this method, a pre-trained VAE \cite{kingma2013auto} compresses images into low-dimensional latent representations. The diffusion and denoising steps then take place within this latent space.

In addition, diffusion models \cite{ho2020denoising, Song2021} have seen broad application in speech synthesis \cite{chen2020wavegrad, kong2020diffwave, leng2022binauralgrad}, image generation \cite{zhang2024mmginpainting, saharia2022photorealistic, xu2024sgdm}, image restoration \cite{huang2024wavedm}, and video generation \cite{liu2023conditional,zhao2024ta2v}. These models transform random noise into high-quality data through a fixed number of steps in a Markov chain process.

\section{DualSpec}
\label{sec:DualSpec}
In this section, we will introduce the proposed DualSpec framework, starting with its workflow pipeline, followed by a detailed explanation of the components it comprises, including the text encoder, diffusion model and VAE.

\begin{figure*}[t]
    \centering
    \includegraphics[width=0.95\textwidth]{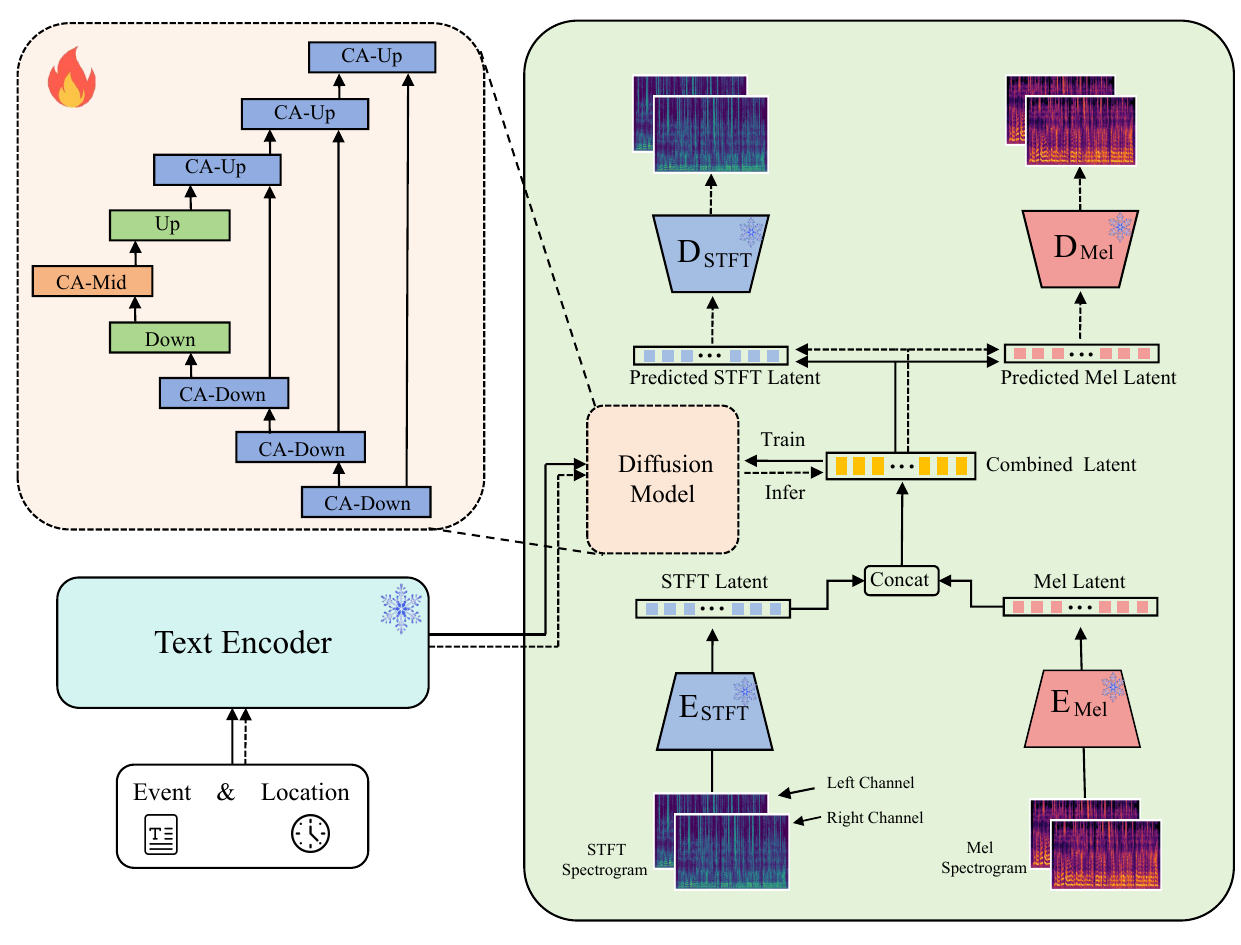}
    \caption{{The pipeline of DualSpec. Solid and dashed lines represent the training phase and inference phase, respectively. ``CA-Down'' and  ``CA-Up'' are the abbreviations for the cross-attention downsampling block and cross-attention upsampling block, respectively. The words ``Down'' and ``Up'' are the abbreviations for the downsampling block and upsampling block, respectively. ``CA-Mid'' is short for the cross-attention mid block.}}
    \label{fig:DualSpec}
\end{figure*}

\subsection{Overview of DualSpec}

The pipeline of the proposed model, DualSpec, is illustrated in Fig.~\ref{fig:DualSpec}. During training, DualSpec first adds multi-level Gaussian noise to the input latent representations, then the diffusion model learns to gradually recover the original features from the noise. Two VAE encoders are used to compress images into latent spaces to reduce computational costs, which separately extract latent representations from Mel spectrograms and STFT spectrograms. These representations are then concatenated and fed into the diffusion model for training. The text encoder maps text prompts into semantic vectors to guide the generation direction. The model is optimized by minimizing the difference between predicted noise and ground truth noise, and a conditional mechanism is employed to achieve precise alignment between text and spatial audio.

During inference, the input text is first converted into semantic features by the text encoder. Starting from random Gaussian noise and utilizing a diffusion model, the process iteratively refines and generates latent representations that align with the textual description. These latent representations are then decoded into Mel and STFT spectrograms through two dedicated VAE decoders. The Mel spectrogram is subsequently inverse-transformed into the amplitude spectrogram \footnote{Convert the Mel spectrogram to a waveform using a vocoder, then perform an STFT transform to extract the amplitude spectrogram.} and merged with the STFT phase spectrogram to create new STFT features. Finally, spatial audio is produced via the inverse STFT (ISTFT) transformation.

\subsection{Text encoder}
{FLAN-T5-Large~\cite{chung2024scaling} is used as the text encoder ($E_{\text{text}}$)}, which generates text encodings $\tau \in \mathbb{R}^{L \times d_{\text{text}}}$, where $L$ is the token length and $d_{\text{text}}$ is the embedding dimension. Through pretraining on chain-of-thought and instructional datasets~\cite{dai2022can}, FLAN-T5 gains the ability to effectively utilize in-context cues and mimic gradient-based optimization through its attention mechanisms. By treating each input as a separate task and leveraging its advanced pretraining, the model excels at extracting task-relevant information while reducing noise, ultimately enabling more accurate conversion from text to acoustic representations.
{We freeze the text encoder, which not only saves computational resources but also effectively prevents overfitting.}


\subsection{Diffusion model for text-guided generation}

LDM~\cite{rombach2022high}  produces the latent representation prior $z_{0}$ under the influence of a text-derived representation $\tau$. This involves approximating the distribution $q(z_{0} | \tau)$ using a trainable model $p_{\theta}(z_{0} | \tau)$. In our method, the model receives two different features: Mel and STFT spectrogram. Specifically, Mel spectrogram is encoded by Mel-VAE, while STFT feature is encoded by STFT-VAE, which produce corresponding latent representations. See Section \ref{sec:DualSpec_vae_arch} for the details of the VAEs. The two latent representations are then concatenated to form a combined latent input, which is processed by a diffusion model. After processing, the combined input is split into two parts, each used for the loss calculation.

LDM accomplishes this task via both forward and reverse diffusion processes. We denote superscripts $m$ and $s$, representing latent diffusion processes for Mel spectrogram and STFT features. The forward process incrementally adds noise to $z^{m}_{0}$ and $z^{s}_{0}$ using a sequence of Gaussian transitions, regulated by noise levels $0 < \beta_{1} < \beta_{2} < \cdots < \beta_{N} < 1$:

\begin{equation}
q(z^{k}_{n} | z^{k}_{n-1}) = \mathcal{N}\left(\sqrt{1-\beta_{n}} \, z^{k}_{n-1}, \beta_{n} \mathbf{I}\right),
\end{equation}
\vspace{-0.5em}

\begin{equation}
q(z^{k}_{n} | z^{k}_{0}) = \mathcal{N}\left(\sqrt{\bar{\alpha}_{n}} \, z^{k}_{0}, (1-\bar{\alpha}_{n}) \mathbf{I}\right),
\end{equation}
\vspace{-0.5em}
\[
k \in \{m,s\},
\]
where \( N \) refers to the total number of diffusion steps, \( \alpha_{n} = 1 - \beta_{n} \), and \( \bar{\alpha}_{n} = \prod_{i=1}^{n} \alpha_{i} \). A reparameterization method~\cite{song2020denoising} simplifies sampling any intermediate states \( z^{m}_{n} \) and \( z^{s}_{n} \) from \( z^{m}_{0} \) and \( z^{s}_{0} \) through the formula:

\begin{equation}
z^{k}_{n} = \sqrt{\bar{\alpha}_{n}} \, z^{k}_{0} + \sqrt{1 - \bar{\alpha}_{n}} \, \epsilon^{k}, k \in \{m,s\},
\end{equation}
where \( \epsilon^{m}, \epsilon^{s} \sim \mathcal{N}(\mathbf{0}, \mathbf{I}) \) introduce independent noise. At the final step of forward diffusion, both \( z^{m}_{N} \) and \( z^{s}_{N} \) resemble standard Gaussians.

In the reverse process, noise is removed to recover \( z^{m}_{0} \) and \( z^{s}_{0} \). The reverse procedure employs a loss function to predict noise for both latents using the text-conditioned model \( \hat{\epsilon}_{\theta} \):

\vspace{-0.5em}
\begin{equation}
\resizebox{0.43\textwidth}{!}{$\mathcal{L}_{DM} = \sum_{k \in \{m,s\}} \sum_{n=1}^{N} \gamma_{n} \mathbb{E}_{\epsilon^{k}_{n} \sim \mathcal{N}(\mathbf{0}, \mathbf{I}), z^{k}_{0}} \left\| \epsilon^{k}_{n} - \hat{\epsilon}_{\theta}^{(n)}(z^{k}_{n}, \tau) \right\|_{2}^{2}$},
\end{equation}
$\hat{\epsilon}_{\theta}$ uses a U-Net structure~\cite{ronneberger2015u} with cross-attention to incorporate text features.

Here, \( \gamma_{n} \) adjusts the weight of each reverse step according to its signal-to-noise ratio. Sampling for \( z^{m}_{n} \) and \( z^{s}_{n} \) follows the previously described formulas, and \( \tau \) represents the text encoding for guidance (see Section~2.1). Noise predictions guide the reconstruction of both latents, modeled as:
\vspace{-0.5em}
\begin{equation}
p_{\theta}(z^{k}_{0:N} | \tau) = p(z^{k}_{N}) \prod_{n=1}^{N} p_{\theta}(z^{k}_{n-1} | z^{k}_{n}, \tau),
\end{equation}
\vspace{-0.5em}
\begin{equation}
p_{\theta}(z^{k}_{n-1} | z^{k}_{n}, \tau) = \mathcal{N}\left(\mu_{\theta}^{(n)}(z^{k}_{n}, \tau), \tilde{\beta}^{(n)}\right),
\end{equation}
\vspace{-0.5em}
\begin{equation}
\mu_{\theta}^{(n)}(z^{k}_{n}, \tau) = \frac{1}{\sqrt{\alpha_{n}}} \left[ z^{k}_{n} - \frac{1 - \alpha_{n}}{\sqrt{1 - \bar{\alpha}_{n}}} \hat{\epsilon}_{\theta}^{(n)}(z^{k}_{n}, \tau) \right],
\end{equation}
\vspace{-0.5em}
\begin{equation}
\tilde{\beta}^{(n)} = \frac{1 - \bar{\alpha}_{n-1}}{1 - \bar{\alpha}_{n}} \beta_{n},
\end{equation}
\vspace{-0.5em}
\[
k \in \{m,s\}.
\]

\subsection{Classifier-free guidance}

For the reverse diffusion process that reconstructs the priors \( z^{m}_{0} \) and \( z^{s}_{0} \), we integrate a classifier-free guidance strategy~\cite{ho2022classifier} conditioned on text input \( \tau \). This approach uses a guidance scale \( w \) during inference to balance text-conditioned and unconditional noise predictions. When text guidance is disabled, we pass an empty input \( \varnothing \), and the guided estimations for Mel and STFT latents are given by:

\vspace{-0.5em}
\begin{equation}
\resizebox{0.43\textwidth}{!}{$\hat{\epsilon}_{\theta}^{(n)}(z^{k}_{n}, \tau) = w \epsilon_{\theta}^{(n)}(z^{k}_{n}, \tau) + (1-w) \epsilon_{\theta}^{(n)}(z^{k}_{n}, \varnothing),k \in \{m,s\}$}.
\end{equation}

\vspace{-0.5em}

\subsection{{Audio VAE}}
\label{sec:DualSpec_vae_arch}
\begin{figure}[t]
    \centering
    \includegraphics[width=0.5\textwidth]{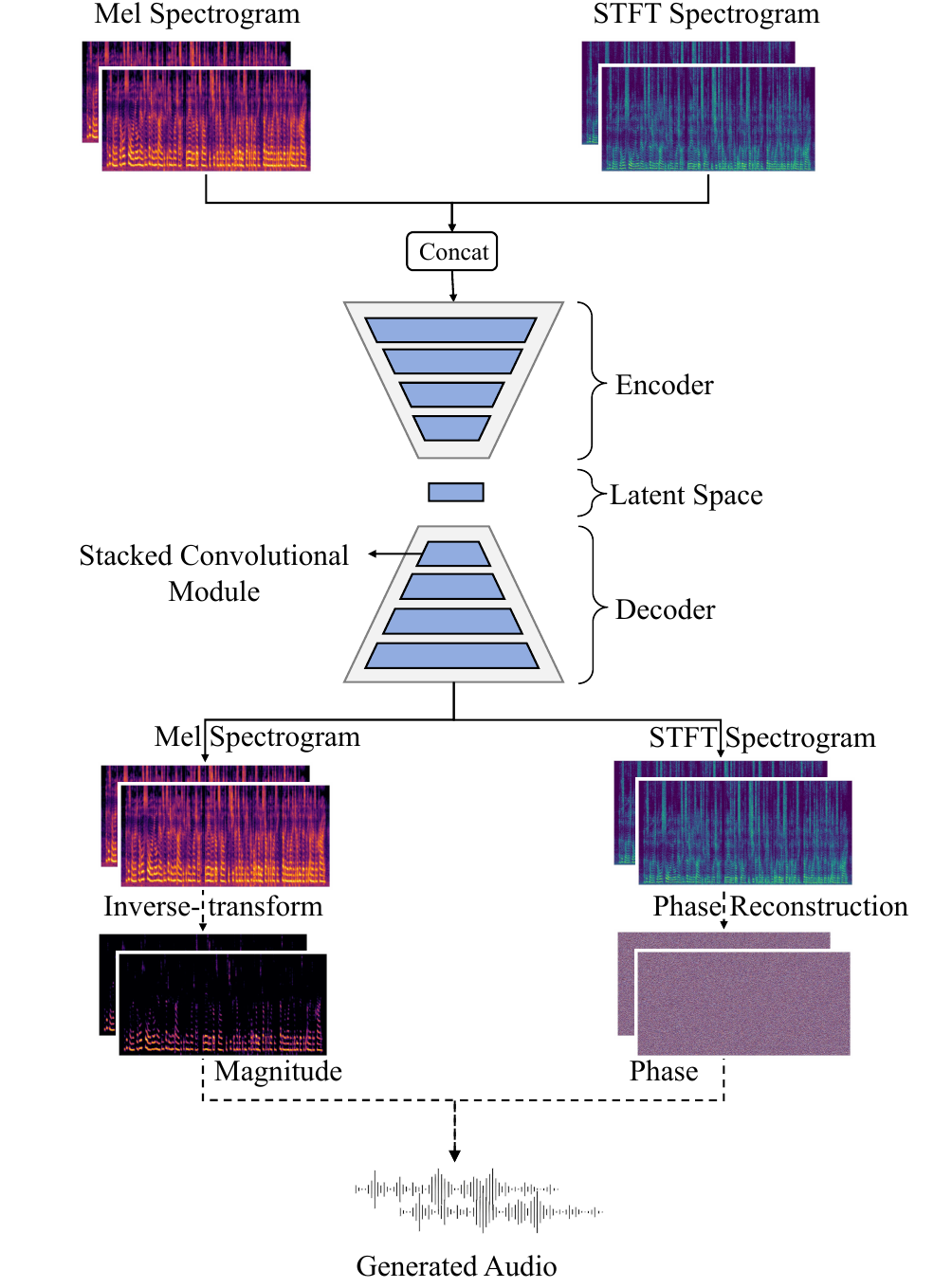}
    \caption{{The structural design of Dual-VAE demonstrates how to integrate STFT and Mel spectrogram features into a unified latent space representation. The dashed line indicates that this process occurs only during the inference phase.}}
    \label{fig:DualSpec_vae}
\end{figure}

The VAE compresses audio features, \( m \in \mathbb{R}^{T \times F} \), into latent representations \( z^{m}_{0} \in \mathbb{R}^{C \times T/r \times F/r} \) and \( z^{s}_{0} \in \mathbb{R}^{C \times T/r \times F/r} \), where \( C \), \( T \), \( F \) and \( r \) denote channel count, time slot count, frequency slot count, and compression ratio, respectively.

The latent diffusion model then uses text guidance \( \tau \) to reconstruct the audio priors \( \hat{z}^{m}_{0} \) and \( \hat{z}^{s}_{0} \). Both the encoder and decoder are built upon stacked convolutional modules \cite{liu2023audioldm} and jointly optimized by maximizing the evidence lower bound (ELBO)~\cite{kingma2013auto} while minimizing adversarial loss~\cite{isola2017image}. We trained two VAE models: Mel-VAE and STFT-VAE, which compress the Mel spectrogram and STFT spectrogram, respectively.

Additionally, in order to explore the diverse combinations of the two acoustic features, we also trained the Dual-VAE model, which takes the concatenation of Mel spectrograms and STFT spectrograms as input and outputs their joint reconstructions, as illustrated in Fig. \ref{fig:DualSpec_vae}\footnote{{
The Inverse-transform in Fig. \ref{fig:DualSpec_vae} is implemented using least squares optimization to minimize the Euclidean distance between the Mel spectrogram and the product of the estimated magnitude spectrogram and the filter banks.
}}. 


\section{Objective evaluation of spatial fidelity}

\label{sec:local}
\begin{figure}[t]
    \centering
    \includegraphics[width=0.5\textwidth]{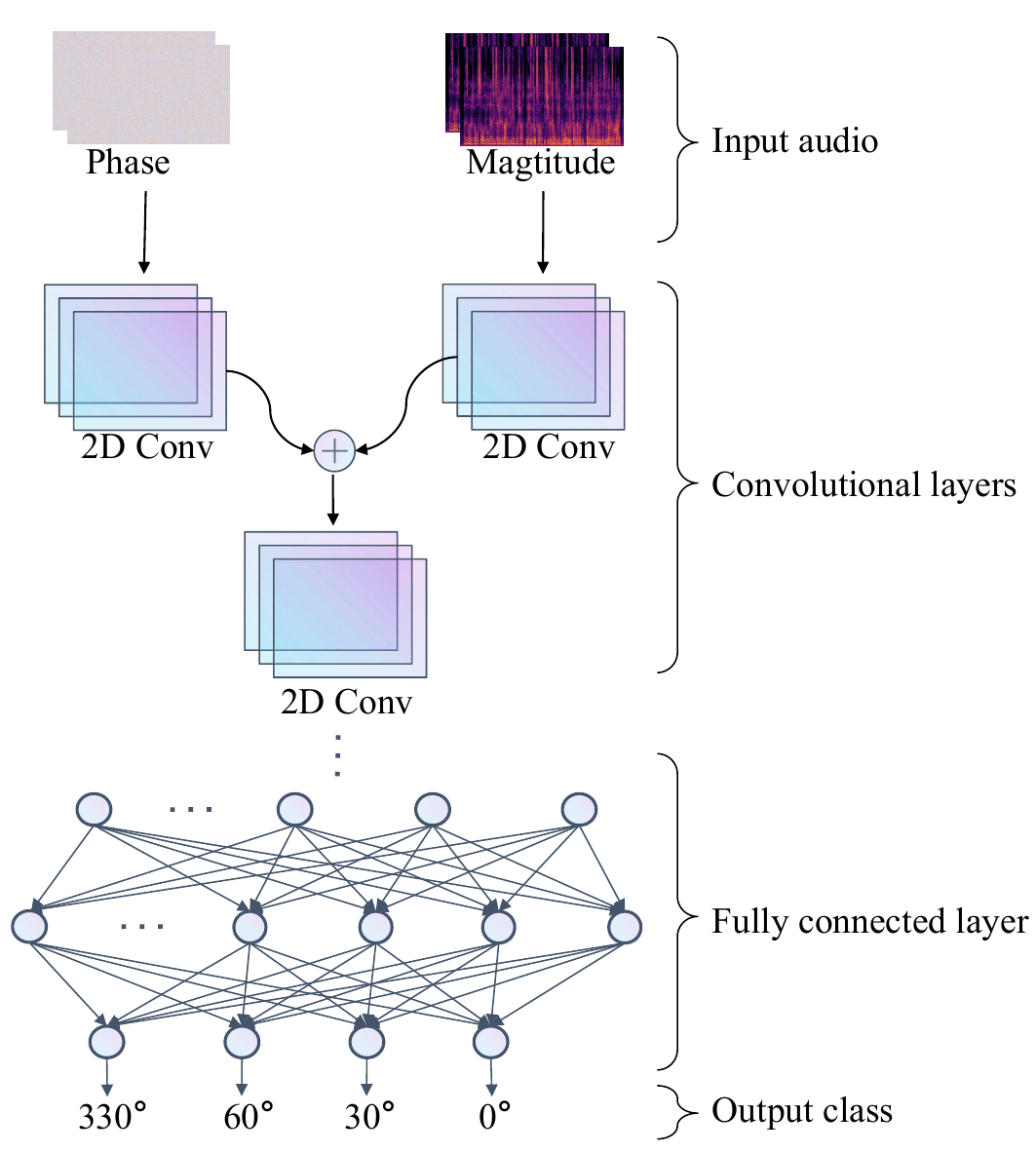}
    \caption{{The structure of the sound localization model.}}
    \label{fig:local}
\end{figure}

    To assess the spatial fidelity of synthesized binaural audio, we employ a DNN-based sound source localization framework \cite{feng2025audiospa} as an objective evaluation metric. When the localization model achieves extremely low errors on ground-truth binaural audio, the predicted DOA for the synthesized binaural audio can be considered as an approximation of its actual DOA. This allows us to compare the predicted DOA with the ground-truth DOA and compute the spatial error.

As shown in Fig.~\ref{fig:local}, we employ a classification-based localization model. When using conventional compact microphone arrays for localization, the process primarily relies on phase spectrograms \cite{feng2025eliminating}, as the amplitude differences between microphones are minimal. However, due to the head shadow effect, the amplitude differences between the two ears can be significant. Leveraging this characteristic, we develop an end-to-end architecture that directly processes phase and magnitude spectrograms as dual-branch inputs.

Building on the phase-based framework from \cite{feng2025eliminating}, we integrate a parallel magnitude spectrogram branch alongside the original phase input, creating a dual-stream architecture that comprehensively utilizes frequency-domain auditory information. By using convolutional layers, our design bypasses manual feature extraction and autonomously learns spatial auditory cues. The first convolutional layer explicitly aligns with binaural physiology through its two-channel structure, enabling more complex nonlinear combinations than interaural time differences and interaural level differences.

As shown in Fig.~\ref{fig:local}, for both input types, we apply three convolutional layers to generate refined feature maps: $\mathbf{E}^{pha'} \in \mathbb{R}^{c' \times f' \times t'}$ (phase features) and $\mathbf{E}^{mag'} \in \mathbb{R}^{c' \times f' \times t'}$ (magnitude features). These parallel-processed features are merged through channel-wise concatenation, creating a composite representation $\mathbf{E}^{com'} \in \mathbb{R}^{2c' \times f' \times t'}$. The combined features then undergo additional convolutional refinement, producing enhanced spatial-temporal features $\mathbf{E}^{com''} \in \mathbb{R}^{c'' \times f'' \times t''}$. This tensor is flattened into a 1D vector for dense processing. Three sequential fully connected layers transform this vector into the final output $\hat{\mathbf{p}} \in [0, 1]^{I}$, representing location probabilities across $I$ azimuth sectors. Our implementation uses 12 azimuth classes ($I=12$) corresponding to angles $\{0^{\circ}, 30^{\circ}, 60^{\circ}, \cdots, 330^{\circ}\}$. Each class spans $30^{\circ}$ intervals, enabling discrete spatial probability estimation across the full circular plane.

\section{Pipeline of synthesizing spatial audio datasets}   \label{sec:data}
In this section, we introduce the monaural sound event datasets, the HRIRs, and the production process of the spatial audio dataset, as well as explain the division of datasets for model training and testing.

\subsection{Monaural sound event datasets}
\label{sec: Sound Event Datasets}

\begin{itemize}
    \item ARCA23K \cite{gemmeke2017audio} is a sound event dataset designed to study real-world label noise, comprising 23,727 audio clips sourced from Freesound. These clips are categorized into 70 classes based on the AudioSet \cite{gemmeke2017audio} classification framework. The dataset was generated through a fully automated pipeline without human review, resulting in a significant proportion of potentially mislabeled audio samples due to the absence of manual quality control.
    \item The UrbanSound8K dataset \cite{salamon2014dataset} is an open-source audio corpus comprising 8,732 labeled short audio clips ($\leq$ 4 seconds) categorized into 10 urban sound classes: \emph{air\_conditioner}, \emph{car\_horn}, \emph{children\_playing}, \emph{dog\_bark}, \emph{drilling}, \emph{engine\_idling}, \emph{gun\_shot}, \emph{jackhammer}, \emph{siren}, and \emph{street\_music}. These classes are derived from the urban sound taxonomy to systematically represent typical urban acoustic scenes. The audio clips, sourced primarily from Freesound, undergo manual verification to ensure labeling accuracy. The dataset serves as a benchmark for environmental sound classification and sound event detection (SED).
    \item The ESC-50 dataset \cite{piczak2015esc} is a collection of 2,000 labeled environmental audio recordings, ideal for benchmarking methods of environmental sound classification. It encompasses 50 distinct categories of sounds sourced from Freesound, including natural, human, and domestic noises. Each category consists of 40 individual 5-second-long audio clips.
    \item The FSD50K dataset \cite{fonseca2021fsd50k} is also a publicly accessible collection of human-annotated sound occurrences, featuring a total of 51,197 audio clips from Freesound that are distributed across 200 categories derived from the AudioSet Ontology \cite{gemmeke2017audio}. It predominantly encompasses sound events generated by physical sources and production mechanisms, including but not limited to human vocalizations, sounds produced by objects, animal vocalizations, natural soundscapes, and musical instrument performances.
\end{itemize}

\subsection{HRIRs}
In this study, we utilized the HUTUBS HRTF dataset \cite{brinkmann2019hutubs}, specifically focusing on the HRIR measurements from one individual identified as `pp96'.
The dataset contains HRIRs captured at 440 distinct locations on a spherical grid. For our specific requirements, which involve only horizontal plane analysis, ignoring vertical variations, we extracted data from 12 positions located precisely at 0° elevation.

\subsection{Data processing pipeline}
\label{subsec: Data Processing}
First, we resampled each mono audio clip in  Section \ref{sec: Sound Event Datasets} to a sample rate of 16kHz, and then we removed all audio clips that were shorter than 1s.
In this way, we obtained a total of more than 90000 mono audio clips. Next, we used HRIRS to convert each mono audio clip into corresponding dual-channel stereo audio files. 
We divided the horizontal plane into 12 directions, each separated by 30 degrees.
Consequently, each mono audio clip was transformed into 12 stereo audio files, resulting in a total of approximately 1.2 million spatial stereo audio files.


For VAE training, we construct each training batch using the spatial audio files mentioned above, cropping the audio length to 5 seconds. If the original audio length is less than 5 seconds, it is zero-padded to ensure the required length is met.

\begin{figure}[t]
    \centering
    \includegraphics[width=0.45\textwidth]{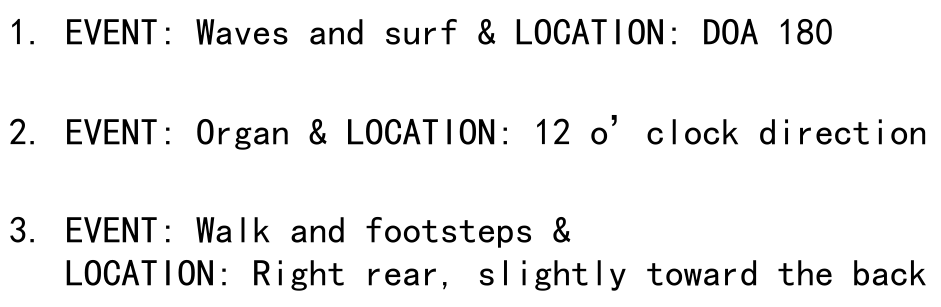}
    \caption{Some examples of creation prompts with sound location descriptors: 1. DOA. 2. Clock. 3. General Description.}
    \label{fig:prompts}
\end{figure}

For the diffusion model training stage, we created a text prompt template that describes both the sound event and its location in the following format: \emph{EVENT: [sound event] \& LOCATION: [sound location]}. We fill the \emph{[sound event]} placeholder with the original class label of the audio. In cases of multiple labels, we randomly select one of them. For \emph{[sound location]}, we used a total of three types of sound location descriptors: DOA, Clock, and General Description. Fig. \ref{fig:prompts} presents some examples of created prompts. We selected 50,000 audio clips from a pool of 1.2 million for training the diffusion model, using the first 5 seconds of each audio clip and padding with zeros if necessary. During training, each prompt randomly adopted one of three location descriptors. 

\subsection{Spatial sound event dataset}
As described in Section 3, we used 1.2 million text-free spatial audio samples to train the VAE in a self-supervised manner. For training the diffusion model, we utilized 50,000 text-audio pairs. To evaluate the performance of both models, we employed a test set of 5,000 spatial audio samples which are distinguished from the training data.

\section{Experiments}
\label{sec:exp}
In this section, we present the implementation details of our method, the evaluation metrics, and the evaluation of VAE and diffusion models.

\subsection{Implementation details of VAEs}
\label{subsec: Implementation details of VAE}
{For the training of the reconstruction of the STFT spectrogram, we configured the VAE with a compression level of 8. We set the number of FFT points and the window length to 512, while the hop length was set to 256. For the reconstruction of the Mel spectrogram, we set the compression level to 4 and extracted 64 bands Mel spectrogram. When training the combination of two features, we followed the aforementioned settings and set the compression level to 8. We utilized the Adam optimizer \cite{kingma2014adam} with a learning rate of $4.5 \times 10^{-6}$ to train all VAE models. The training dataset, as described in Section \ref{subsec: Data Processing}, consisted of 1.2 million samples. }

{We trained three VAEs with different acoustic features: Mel-VAE, STFT-VAE and Dual-VAE. As their names suggest, Mel-VAE and STFT-VAE were trained on Mel spectrograms and STFT spectrograms, respectively. The Dual-VAE, as described in Section \ref{sec:DualSpec_vae_arch}, was trained on both types of acoustic features simultaneously}. {
All VAE models were trained on 16 NVIDIA H100 GPUs, with approximately 0.6 million training steps for Mel-VAE, 1.2 million for STFT-VAE, and 2.4 million for Dual-VAE.
The batch sizes per GPU were set to 16, 8, and 4 for Mel-VAE, STFT-VAE, and Dual-VAE, respectively. To stabilize the training process, adversarial loss was not applied during the first 60 thousand steps.}

\subsection{Implementation details of diffusion models}
\label{subsec: Implementation details of diffusion models}

{Our diffusion model was built upon the UNet architecture from Stable Diffusion \cite{rombach2022high}. In this model, we configured it with 8 channels and a cross-attention dimension of 1024. Pre-trained weights \cite{majumder2024tango}, which were trained on the AudioCaps mono audio dataset \cite{kim2019audiocaps}, were used. We used the AdamW optimizer \cite{loshchilov2017fixing} with a learning rate of $3 \times 10^{-5}$ and a weight decay of $1 \times 10^{-2}$. Additionally, a linear learning rate scheduler \cite{li2019budgeted} was employed to adjust the learning rate over the course of training.}
We trained the diffusion model for 200 epochs, using 4 NVIDIA H100 GPUs, with each GPU having a batch size of 2. Gradient accumulation was set to 8, resulting in an effective total batch size of 64. We set the number of inference steps to 200 and the classifer-free guidance to 3. {We employed the pre-trained FLAN-T5-Large~\cite{chung2024scaling} model as the text encoder and keep its parameters frozen during the training process.}

We have designed five different TTSA models, namely Mel-base, STFT-base, DualSpec, DualSpec-D, and DualSpec-P. Among them, Mel-base and STFT-base denoted diffusion models trained with Mel features and STFT features, respectively. They utilized Mel-VAE and STFT-VAE to compress acoustic features, respectively. DualSpec trained the diffusion model using a combination of latent representations encoded by the Mel-VAE and STFT-VAE, which is illustrated in Fig. \ref{fig:DualSpec}. DualSpec-D utilized the latent representations encoded by Dual-VAE in Fig. \ref{fig:DualSpec_vae} to train the diffusion model. DualSpec-P adopted an architecture similar to DualSpec, but utilized the VAE in \cite{liu2023audioldm} to encode Mel feature. Notably, this VAE can only process single-channel audio. Therefore, during the training process, we used VAE to encode Mel features on each channel of the spatial audio separately, and then concatenated the latent representations. 
{
For Mel-base, DualSpec, and DualSpec-P, we used the HiFi-GAN vocoder \cite{kong2020hifi} with pre-trained weights from \cite{liu2023audioldm} to convert the Mel spectrogram into an audible waveform.}

\subsection{Evaluation metrics}

\subsubsection{Generation Quality Evaluation}
\label{rqe}
{We employ the following metrics to evaluate the generation quality of our VAEs and diffusion models:}

\begin{itemize}
    \item { \textit{Peak Signal-to-Noise Ratio (PSNR)}: It reflects the extent of signal distortion by quantifying the difference between the original and test signal. Its mathematical definition is as follows:
    \begin{equation}
        \text{PSNR} = 10 \cdot \log_{10} \left( \frac{\text{MAX}^2}{\text{MSE}} \right),
    \end{equation}
    where MAX represents the maximum possible value of the audio signal, and MSE denotes the mean squared error between the original audio and the test audio.} 
    \item {\textit{Structural Similarity Index (SSIM)}: It evaluates the degree of similarity by analyzing the correlations between signals in terms of luminance, contrast, and structure. Specifically, it is implemented by comparing the mean, variance, and covariance of the signals, with its definition as follows:
    \begin{equation}
         \text{SSIM} = \frac{(2\mu_x\mu_y + c_1)(2\sigma_{xy} + c_2)}{(\mu_x^2 + \mu_y^2 + c_1)(\sigma_x^2 + \sigma_y^2 + c_2)},
    \end{equation}    
    where $u_x$ and $u_y$ represent the mean values of the test signal and the reference signal, respectively. $\sigma_x^2$ and $\sigma_y^2$ are their corresponding variances, and $\sigma_{xy}$ represents the covariance between the two signals. $c_1$ and $c_2$ are constant parameters related to the dynamic range of the audio.}
    
    \item {\textit{Frechet Distance (FD)}: It is a metric commonly used to assess the quality of generated audio by quantifying the discrepancy between the feature distributions of real and generated audio. Specifically, audio signals are first encoded into feature vectors using the PANNs audio feature extraction model \cite{kong2020panns}, after which the FD score is computed according to the following formula:
    \begin{equation}
        \text{FD} = \|\bar{\mu}_y - \bar{\mu}_x\|^2 + \text{Tr}\left(\Sigma_y + \Sigma_x - 2 \cdot \sqrt{\Sigma_y \Sigma_x}\right),
    \end{equation}  
    where $\bar{\mu}_x$ and $\bar{\mu}_y$ denote the mean vectors of the generated and real audio features across all dimensions, respectively, while $\Sigma_x$ and $\Sigma_y$ represent the corresponding covariance matrices.}

    \item {\textit{Kullback-Leibler Divergence (KL)}: It is used to quantify the discrepancy between the probability distributions of real and generated audio. Before computing the divergence, audio features are extracted using the PANNs model. A smaller KL divergence indicates that the distribution of the generated audio is closer to that of the real audio.}
    \item {\textit{Inception Score (IS)}: It assesses both the quality and diversity of generated samples, and its calculation is defined as follows:
    \begin{equation}
        \text{IS} = \exp \left( \mathbb{E}_x \left[ \text{KL}(p(y|x) \parallel p(y)) \right] \right),
    \end{equation}  
    where $p(y|x)$ represents the classification probability distribution for a single audio sample, $p(y)$ denotes the marginal distribution over all samples, and $\text{KL}(A \parallel B)$ indicates the KL score between distributions A and B. The pre-trained classification model is PANNs.}
\end{itemize}

\subsubsection{Spatial Perception Evaluation}
For spatial performance assessment, we employ standard sound source localization metrics from \cite{feng2025eliminating}, which are detailed as follows:

\begin{itemize}
    \item { \textit{Classification Accuracy (ACC)}: 
    It measures the proportion of generated audio samples whose spatial directions are consistent with those of the corresponding real audio.
    The training data includes audio from 12 different DOAs, with audio from each DOA categorized into a separate class. We use the sound source localization model from Section IV to obtain the DOA labels for the generated audio. The ACC is formulated as follows:
    \begin{equation}
    \text{ACC} = \frac{N_{\text{correct}}}{N_{\text{total}}},
    \end{equation}
    where $N_{\text{correct}}$ denotes the number of generated audio clips whose DOA matches that of the corresponding real audio clips, and $N_{\text{total}}$ denotes the total number of generated audio clips.}

    \item {\textit{Mean Absolute Error (MAE)}: It is a commonly used metric for evaluating the performance of direction estimation models. Due to the periodic nature of directional angle values, MAE is adopted as an appropriate metric to quantify the difference between the DOAs of the generated and real audio, thereby improving the reliability of the evaluation. Given a ground-truth angle $\theta$ and its corresponding prediction $\hat{\theta}$ for a sample, the MAE computation follows this angular distance formulation:
    \begin{equation}\label{eq:mae}
        \mathrm{MAE}(\circ)= \mathrm{min}(|\hat \theta - \theta|_1, 360-|\hat \theta - \theta|_1).
    \end{equation}
    }
\end{itemize}

\subsection{Evaluation of VAEs using different features}
\label{subsec: Evaluation of VAEs Using Different Features}

\begin{table*}[t]
    \centering
    \caption{Evaluation results for different VAE reconstruction performances. {$\downarrow$ indicates that lower values are better, and vice versa for $\uparrow$.}}
    \label{tab:VAE performance}
    \setlength{\tabcolsep}{6.3mm}
    \begin{tabular}{cccccccc}
        \toprule
        \textbf{Model} & \textbf{PSNR$\uparrow$} & \textbf{SSIM$\uparrow$} & \textbf{FD$\downarrow$} & \textbf{KL$\downarrow$} & \textbf{IS$\uparrow$} & \textbf{MAE$\downarrow$} & \textbf{ACC$\uparrow$} \\
        \midrule
        Mel-VAE & 25.42 & 0.841 & 10.42 & 0.590 & 11.18 & 27.93 & 30.02 \\
        STFT-VAE & \textbf{31.93} & \textbf{0.929} & \textbf{10.37} & \textbf{0.461} & \textbf{12.94} & \textbf{5.32} & \textbf{95.50} \\
        Dual-VAE & 27.88 & 0.853 & 10.66 & 1.031 & 10.86 & 7.15 & 93.32 \\
        \bottomrule
    \end{tabular}
\end{table*}

Table \ref{tab:VAE performance} presents the evaluation metrics for three VAEs, including Mel-VAE, STFT-VAE, and Dual-VAE, in terms of their generation performance.

{First, the performance of the three types of VAEs on audio quality generation is presented. The STFT achieves the best performance in the generation quality, attaining the highest scores of 31.93 in PSNR, 0.929 in SSIM, 10.37 in FD, 0.461 in KL divergence and 12.94 in IS. Although Mel-VAE shows competitive results with the SSIM, FD, KL, and IS scores, it exhibits a lower PSNR score of 25.42 than STFT-VAE's 31.93. This indicates that, in terms of audio quality generation based on VAE, the STFT spectrogram outperforms the Mel spectrogram. This conclusion stands in sharp contrast to the generation performance of diffusion models, as will be further confirmed by subsequent experiments. The overall performance of Dual-VAE is not as good as that of STFT-VAE and Mel-VAE due to the loss of information during the inverse transformation from the Mel spectrogram to the STFT magnitude spectrogram. However, it has shown moderate performance in terms of PSNR and SSIM metrics. }

{Next, we present the performance of different VAE models in terms of the direction accuracy of the generated audio. Compared to the other two methods, the STFT spectrogram can also provide more precise spatial position generation, possessing the best MAE and ACC scores. The superiority stems from the fact that the phase spectrogram of the STFT is rich in audio location information, which is lacking in the Mel spectrogram. Therefore, the Mel-VAE have the lowest MAE and ACC scores. Although DualSpec does not match the first two in terms of audio quality generation, it nearly reaches the level of STFT in spatial position generation and significantly outperforms Mel-VAE.}

\subsection{Comparison of different generation methods}

\begin{table*}[ht]
\centering
\caption{\textbf{Performance comparison of the proposed different models with baselines.}}
\setlength{\tabcolsep}{2.5mm}
\begin{tabular}{cc>{}c>{}c>{}cccccc}
\toprule
& \textbf{Model}  & \textbf{Params(M)} & \textbf{FLOPs(G)} & \textbf{Inference Time(s)} & \textbf{FD$\downarrow$} & \textbf{KL$\downarrow$} & \textbf{IS$\uparrow$} & \textbf{MAE$\downarrow$} & \textbf{ACC$\uparrow$} \\
\midrule
\multirow{4}{*}{\textbf{{Baseline}}}
& AudioLDM \cite{liu2023audioldm} & 809 & 452 & 0.36 & 21.45 & 2.470 & 7.194 & - & -  \\
& AudioLDM2 \cite{liu2024audioldm} & 1034 & 537 & 0.71 & 29.19 & 2.425 & 8.554 & - & -   \\
& TANGO \cite{ghosal2023tango} &  1317 & 646 & 0.53 & 27.40 & 2.919 & 9.572& -& -  \\
& TANGO2 \cite{majumder2024tango} & 1317 & 646 & 0.52 & 31.64 & 2.910 & 9.835 & - & -  \\
\midrule
\multirow{5}{*}{\textbf{Ours}}
& Mel-base & 1485 & 679 & 0.65 & \textbf{15.47} & \textbf{2.058} & \textbf{10.691} & 32.28 & 24.54 \\
& STFT-base &1262 & 848 & 0.97& 24.72 & 2.388 & 7.578 & \textbf{9.56} & \textbf{84.78} \\
& DualSpec &1541 & 1145 & 1.21& 23.87 & 2.376 & 7.590 & 13.16 & 79.02 \\
& DualSpec-D & 1262 & 887 & 1.06& 21.70 & 2.384 & 6.152 & 17.32 & 76.01 \\
& DualSpec-P &1373 & 1417 & 1.82& 20.49 & 2.322 & 10.571 & 10.79 & 83.86 \\
\bottomrule
\end{tabular}
\label{tab:diff_comp}
\end{table*}
\begin{figure*}[t]
    \centering
    \includegraphics[width=0.85\textwidth]{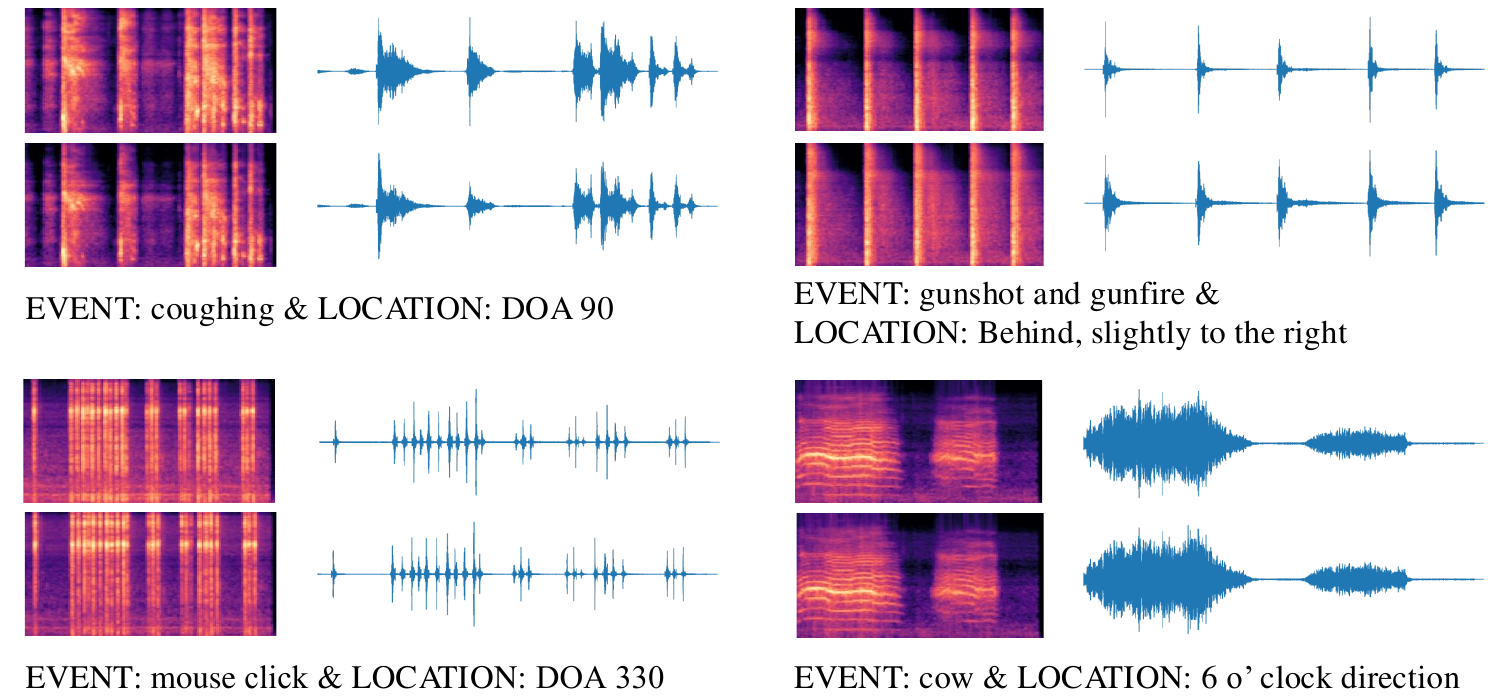}
    \caption{Examples generated by DualSpec-P. The upper section of each example represents the left channel, while the lower section corresponds to the right channel.}
    \label{fig:demo}
\end{figure*}

Table \ref{tab:diff_comp} presents a comparison of the proposed different diffusion models against baseline models across multiple performance metrics. Given the current absence of comparable spatial audio generation works with single-channel reference \footnote{The current text-guided spatial audio generation methods depend on mono reference audio as input \cite{feng2025audiospa}. Strictly speaking, it cannot be qualified as a true TTSA method and thus will not be compared in this context.}, our comparison is limited to recent monaural audio generation approaches, specifically AudioLDM \cite{liu2023audioldm}, AudioLDM2 \cite{liu2024audioldm}, TANGO \cite{ghosal2023tango}, and TANGO2 \cite{majumder2024tango}.

As shown in the table, our proposed method significantly outperforms all baseline methods, particularly in terms of the FD and IS metrics. Notably, even the best-performing baseline method, AudioLDM, falls significantly short compared to Mel-base and DualSpec-P.

{As expected, Mel-base performs best in terms of generation quality, while STFT-base leads in azimuth accuracy. Although the STFT-based generated audio exhibits high directional consistency with text descrption, its audio quality is significantly inferior to that of Mel-based methods. DualSpec achieves higher audio quality by simultaneously leveraging the encoded latent information in both Mel spectrograms and STFT spectrograms compared to using STFT alone. Meanwhile, it nearly matches the localization accuracy of methods based on STFT. Compared to DualSpec, DualSpec-D performs excellently in FD scores, but falls short in the IS metric.}

Compared to DualSpec, DualSpec-P performs better in all evaluation metrics, mainly due to the advantages of training the VAE on the large-scale AudioSet dataset \cite{gemmeke2017audio}, which significantly reduces information loss during the extraction of latent features from Mel spectrograms. DualSpec-P can be considered as a trade-off between Mel-base and STFT-base methods. Fig. \ref{fig:demo} shows several examples generated by this method. Compared to DualSpec, DualSpec-P has achieved a significant improvement in directional accuracy. This indicates that the signal amplitude retains a certain amount of positional information, which is primarily determined by the intensity difference between the left and right channels. 

{
Table \ref{tab:diff_comp} also presents a comparison of the computational load across models, including parameter sizes, FLOPs for one second of audio, and inference time for the same duration.
Compared to the baseline models, DualSpec and its variants show a relatively significant increase in both FLOPs and inference time. The main reasons are as follows. First, due to the use of two features rather than relying solely on the Mel spectrogram as in the baseline method, DualSpec incurs higher computational costs. Second, spatial audio contains more data than ordinary mono audio, which also increases computational costs to some extent.}

\subsection{Ablation study}
\label{subsec: Ablations on Direction Caption}

\subsubsection{Effect of different location descriptors}
\label{subsubsec: Effect of different location descriptors}

\begin{table*}[ht]
\centering
\caption{Evaluation results for different location descriptors.}
\setlength{\tabcolsep}{6.7mm}
\begin{tabular}{c c c c c c c}
\toprule
\textbf{Model} & \textbf{location descriptor} & \textbf{FD$\downarrow$} & \textbf{KL$\downarrow$} & \textbf{IS$\uparrow$} & \textbf{MAE$\downarrow$} & \textbf{ACC$\uparrow$} \\
\midrule
\multirow{3}{*}{Mel-base} & DOA & 15.81 & \textbf{2.032} & 10.51 & \textbf{30.64} & 25.80 \\
& Clock & \textbf{15.42} & 2.056 & \textbf{10.54} & 31.36 & \textbf{26.46} \\
& General Description & 15.67 & 2.036 & 10.17 & 37.37 & 20.40 \\
\midrule
\multirow{3}{*}{STFT-base} & DOA & 24.45 & 2.402 & 7.75 & \textbf{6.02} & \textbf{94.34} \\
& Clock & \textbf{24.41} & 2.385 & \textbf{7.82} & 6.75 & 94.14 \\
& General Description & 24.65 & \textbf{2.389} & 7.55 & 15.43 & 64.68 \\
\midrule
\multirow{3}{*}{DualSpec} & DOA & \textbf{23.79} & 2.371 & 7.28 & 10.61 & 87.92 \\
& Clock & 23.88 & 2.391 & 7.29 & \textbf{9.86} & \textbf{88.04} \\
& General Description & 24.18 & \textbf{2.341} & \textbf{7.33} & 19.70 & 60.76 \\
\midrule
\multirow{3}{*}{DualSpec-D} & DOA & 21.81 & 2.370 & 5.98 & 14.95 & 83.24 \\
& Clock & \textbf{21.70} & 2.387 & \textbf{6.02} & \textbf{14.57} & \textbf{84.10} \\
& General Description & 22.09 & \textbf{2.368} & 5.95 & 23.10 & 59.28 \\
\midrule
\multirow{3}{*}{DualSpec-P} & DOA & 20.84 & 2.309 & \textbf{10.53} & 7.83 & 92.92 \\
& Clock & 20.86 &  \textbf{2.290} & 10.52 & \textbf{7.72} & \textbf{93.17} \\
& General Description & \textbf{20.78} & 2.306 & 10.32 & 17.12 & 63.81 \\
\bottomrule
\end{tabular}
\label{tab:abla}
\end{table*}


Table \ref{tab:abla} illustrates the impact of three different location descriptors on the performance across five proposed models. For each descriptor, the evaluation dataset corresponds to the same sound events as those in the 5000-sample standard test set (refer to Section \ref{subsec: Data Processing}). It is observed that despite variations in location descriptors, the evaluation results for generation quality metrics remain quite similar across models. However, the accuracy of spatial position in the generated audio shows significant variation with changes in these descriptors. Specifically, using DOA and Clock leads to more precise spatial stereo sound generation compared to a General Description. Given that descriptors such as DOA and Clock inherently provide more detailed spatial location information than the General Description, this result is actually as expected.

\subsubsection{Effect of classifier-free guidance and inference steps}
\label{subsubsec: Effect of classifier-free guidance and inference steps}
\begin{figure}[t]
    \centering
    \includegraphics[width=0.4\textwidth]{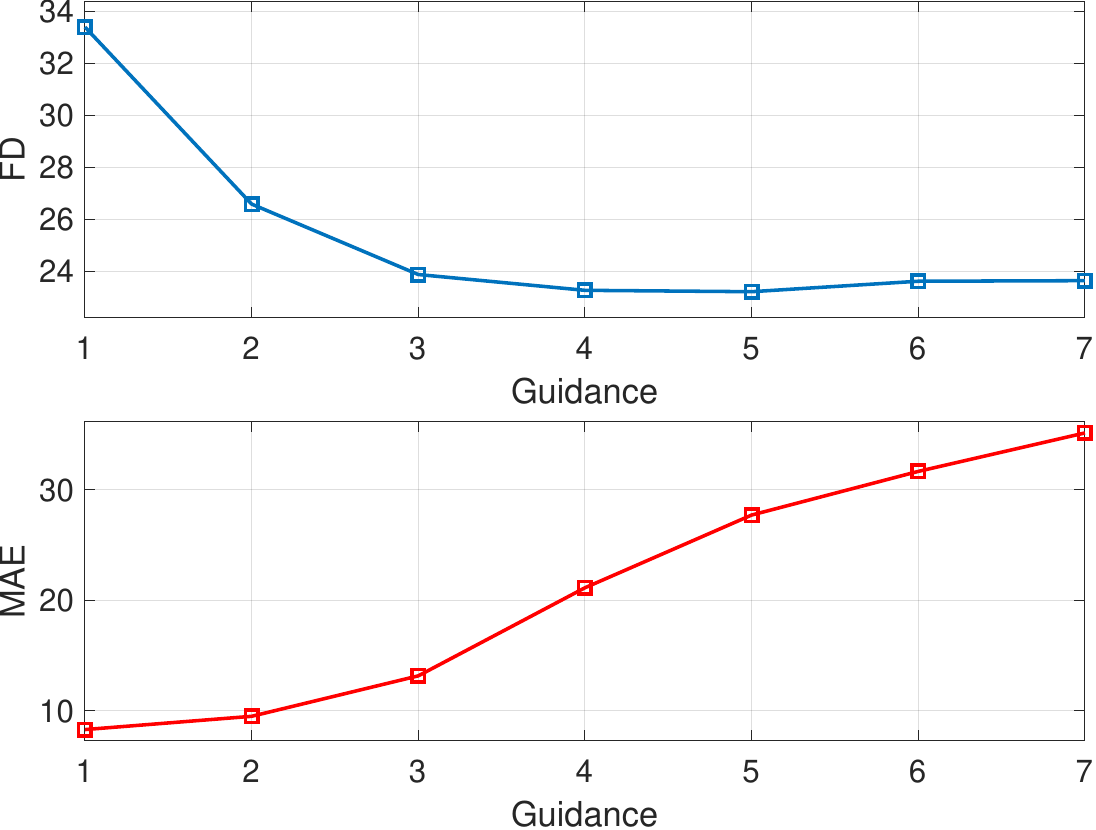}
    \caption{Effect of different classifier-free guidance on the performance of the DualSpec.}
    \label{fig:guide}
\end{figure}

{Fig. \ref{fig:guide} illustrates the effect of different classifier-free guidance on the performance of the DualSpec. As the guidance value increases, the audio generation quality of DualSpec gradually decreases, a trend that is reflected in the FD scores. When the classifier-free guidance reaches 3, this trend begins to flatten. However, the azimuth accuracy of the model shows an opposite trend. As the guidance value increases, the DualSpec's MAE also rises, indicating that a higher classifier-free guidance leads to greater directional errors. To balance both aspects, we set the guidance value to 3.}

\begin{figure}[t]
    \centering
    \includegraphics[width=0.4\textwidth]{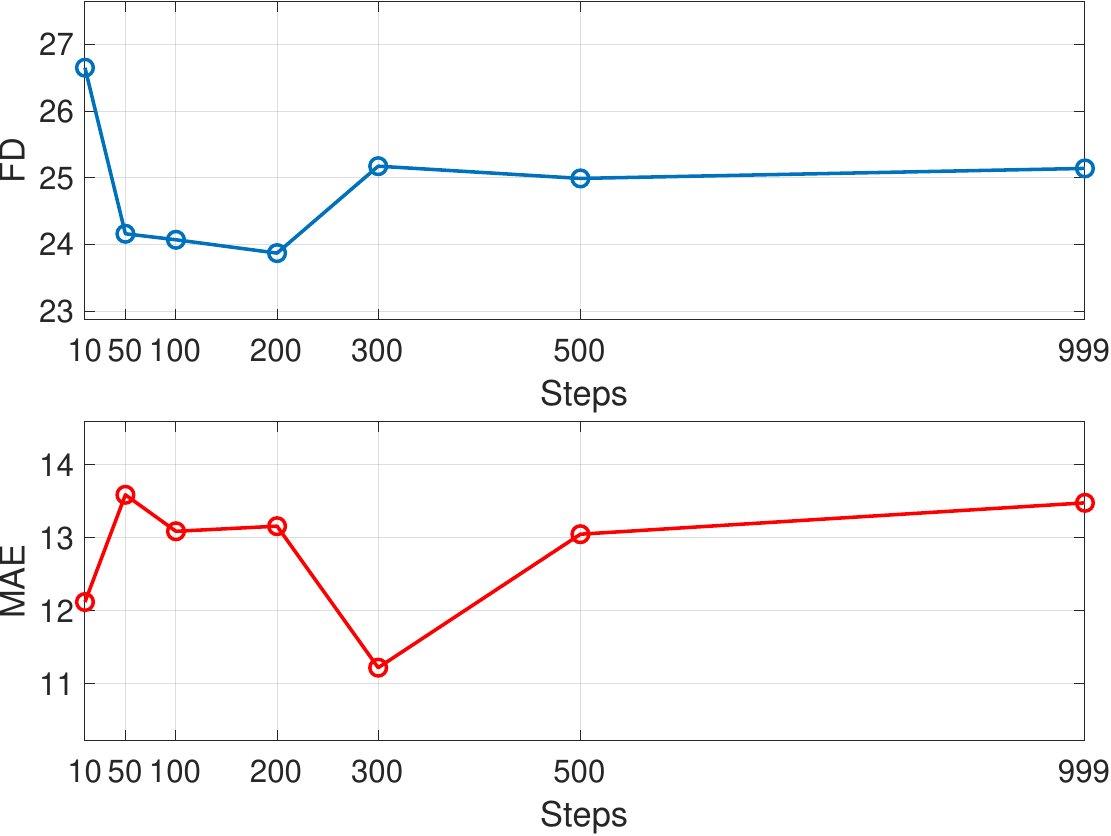}
    \caption{Effect of different number of inference steps on the performance of the DualSpec.}
    \label{fig:step}
\end{figure}

{Fig. \ref{fig:step} illustrates the effect of different number of inference steps on the performance of the DualSpec. We can observe that the FD and MAE scores of DualSpec generally first decrease and then increase as the number of inference steps increases. Specifically, the optimal value for FD is found at the 200th step, whereas the optimal value for MAE is attained at the 300th step. Considering that the angular error between the two step settings is not significantly different, we have chosen the setting with better audio quality, which is to set the number of inference steps to 200.}

\section{Conclusions}
\label{sec:con}

In conclusion, the proposed DualSpec demonstrates significant potential in TTSA generation, offering a novel approach that eliminates the need for monaural audio inputs. 
{
In addition, DualSpec effectively mitigates the limitations inherent in relying on a single feature, as the Mel spectrogram exhibits low azimuth accuracy and the STFT spectrogram leads to degraded audio quality.
We also introduce spatial perception metrics to evaluate the azimuth accuracy of the generated spatial audio.} 
Furthermore, the pipeline for the spatial audio dataset production also lays the foundation for TTSA tasks.
{Experimental results demonstrate that the proposed method can generate spatial audio that balances both audio quality and directional consistency.}

\small
\bibliographystyle{IEEEtran}
\bibliography{Reference}

\end{document}